\documentclass[article,twocolumn,showpacs,preprintnumbers,amsmath,amssymb]{revtex4}
\usepackage{graphicx}% Include figure files
\usepackage{dcolumn}% Align table columns on decimal point
\usepackage{bm}% bold math

\begin{document}
%
%\newpage
%\begin{CJK*}{GBK}{kai}

%\preprint{APS/123-QED}

\title{Spin density wave in oxypnictide superconductors in a three-band model}
\author{Mingsheng Long, Liangbin Hu, W. LiMing}\email{Corresponding author: wliming@scnu.edu.cn}

\affiliation{Dept. of Physics, and Institute for Condensed Matter Physics, School of
Physics and Telecommunication Engineering,  South China Normal University, Guangzhou
510006, China}

\date{\today}
 %\keywords{spin density wave, iron-based oxypnictide
%superconductors, spin susceptibility}

\begin{abstract}
The spin density wave and its temperature dependence in oxypnictide are studied in a
three-band model. The spin susceptibilities with various interactions are calculated in
the random phase approximation(PPA). It is found that the spin susceptibility peaks
around the M point show a spin density wave(SDW) with momentum (0, $\pi$) and a clear
stripe-like spin configuration. The intra-band Coulomb repulsion enhances remarkably the
SDW but the Hund's coupling weakens it. It is shown that  a new resonance appears at
higher temperatures at the $\Gamma$ point indicating the formation of a paramagnetic
phase. There is a clear transition from the SDW phase to the paramagnetic phase.
\end{abstract}

\maketitle
\section{INTRODUCTION}
The high temperature superconductivity in the newly discovered oxypnictides, LnFeAsO (Ln
= La,Pr,Ce,Sm ), has attracted great attention aiming to identify the mechanism of
superconductivity in these materials\cite{Kamihara}. Recently the transition temperature
$T_c$ is dramatically raised from $26K$ to $43K$\cite{Chen}. In addition to the high
$T_c$, these materials display many other interesting properties. The most interesting
phenomena is the presence of competition between the magnetically ordered ground
states\cite{dong} of spin density wave(SDW) and superconductivity, but a controversy
model explains the superconductivity by means of antiferromagnetic spin fluctuation in
LaFeAsO\cite{Mazin}. Pure oxypnictide is not superconducting but shows an anomaly at
about $150 K$ in both resistivity and dc magnetic susceptibility\cite{Kamihara,Ma}. Both
experimental and theoretical evidences show that the anomaly is caused by the SDW
instability\cite{dong, Chen1, McGuire}.

%   The high transition temperatures and the electronic structure of
%the iron-based pnictide superconductors suggest that the pairing
%interaction is of electronic origin\cite{Boeri}.
It is shown by the first principle calculations that the band structure of LnFeAsO near
the Fermi surface (FS) is formed by a  hole-like pocket centered around the $\Gamma$
point and an electron-like pocket around the M point in the extended Brillouin
zone(BZ)(one Fe atom per unit cell)\cite{Singh,Boeri,Haule}. A strong FS nesting effect
exists between the hole and electron pockets with commensurate wave vectors, $(\pi, 0)$
and its symmetric ones. This leads to a strong SDW instability, and is believed to cause
the the anomaly at $150 K$.

Raghu {\it et al} calculated the SDW in the iron oxypnictides within a minimal two-band
model in the phase random approximation(RPA). They found that the SDW is enhanced
significantly by the intra-band Coulomb repulsion. The influence of the inter-band
Coulomb interaction and the Hund's coupling, however, have not yet been fully studied in
literatures. Different researchers select different groups of interaction parameters, but
the relation between them has not been revealed. In addition, the temperature dependence
of the SDW has hardly been theoretically considered. As pointed out by Lee and
Wen\cite{Lee}, a three-band model reproduces more accurately the band structure near the
Fermi surface of oxypnictides. In this work we study the SDW in a three-band model  in
the RPA and the temperature dependence of the spin susceptibility. The magnetic
instability is studied for a wide range of interaction parameters.  We found a new
resonance of the spin susceptibility around the $\Gamma$ point at higher temperatures,
indicating the formation of a paramagnetic phase. There appears a transition from the SDW
phase to the paramagnetic phase when temperature increases.

%The spontaneously symmetry-broken SDW state is characterized in terms of reduce carrier
%density due to (partial) FS nesting, enhanced conductivity duo to the reduction of the
%scattering channel, and loss of 4-fold rotational symmetry with negligible change of
%lattice \cite{dong}.

%\begin{figure}
%\includegraphics[width=8cm,height=6cm]{orbital3.eps}
%\caption{(a)The $Fe$ ions form a square lattice,the crystallographic unit cell contain
%two $Fe$ ions and two $As$ ions .The $As$ ions are not  in the plane of the $Fe$ ions,
%which are somewhat above (solid circles) or below (dashed circles) the plane. (b) A
%schematic diagram showing the hopping parameters between $d_{xz}$ and $d_{yz}$ orbitals.}
%\end{figure}

\begin{figure}\label{hop}
\includegraphics[width=6cm,height=5cm]{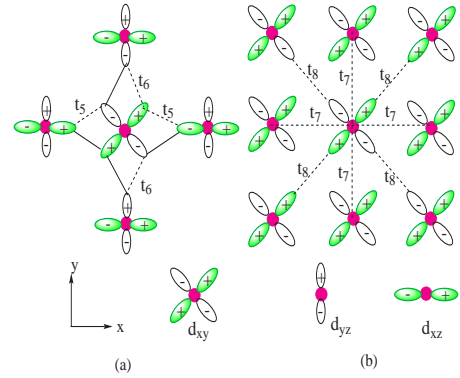}
\caption{(a) A schematic diagram showing the hopping parameters between $d_{xz}(d_{yz})$
and $d_{xy}$. (b) The hopping parameters between the nearest and next nearest $d_{xy}$
orbitals. }
\end{figure}

%\begin{figure}
%\includegraphics[width=6cm,height=8cm]{orbitals322.eps}
%\caption{A schematic showing the hopping parameters of the orbitals
 %between $d_{xz}(d_{yz})$ and $d_{xy}$. for them have different symmetry,
% we only considering the nearest neighbor hopping parameter $t_5,t_6$ between
% $d_{xz}$ and $d_{xy}$ and between $d_{yz}$ and $d_{xy}$ respectively.}
%\end{figure}

\section{Model Hamiltonian}
The FeAs layer in LaFeAsO forms a square lattice, where Fe ions locate on the lattice
sites and an As ion sits at the center of each square. Various band structure
calculations showed that the main contribution to the density of states near the FS comes
from $d_{xz},d_{yz}$ and $d_{xy}$ of the five $3d$ orbitals of Fe atoms\cite{Lee}. The
left two orbitals are far apart from the FS.

%Here we consider a two-dimensional lattice with three degenerate $d_{xz}$, $d_{yz}$ and
%$d_{xy}$ orbitals per site. The tight-binding coupling between the three orbital ions are
%illustrated in Fig.1(b). Here $t_1, t_2$ are the nearest-neighbor hopping parameters
%between $d_{xz}$ orbitals, and $d_{yz}$ ones. $t_3,t_4$ are the corresponding
%next-nearest-neighbor hopping parameters. Secondly, we consider the nearest-neighbor
%admixture between $d_{xz}(d_{yz})$ and $d_{xy}$ along the $y(x)$-axis. These orbitals
%couples because of the asymmetry induced by the $As$ ions.
%This implies that the $d_{xz},
%d_{xy}$ overlap integral is odd under $x\rightarrow-x$ and thus vanishes. Only the
%$d_{yz}, d_{xy}$ matrix element survives\cite{Lee}. This allows us to conclude that the
%nearest neighbor $d_{xz}, d_{xy}$ and $d_{yz}, d_{xy}$ mixing give rise to
%$\varepsilon_{13}=-\varepsilon_{31}=-2 i t_5 \sin k_x \sin k_y $ and $
%\varepsilon_{23}=-\varepsilon_{32}=-2 i t_6\sin k_x \sin k_y$ \cite{Lee}. We only
%consider the nearest hopping because they have different symmetry in X-Y plane, the
%hopping is weak.

In the three band model, the hopping terms between $d_{xz}$, $d_{yz}$ and $d_{xy}$ as
illustrated in Fig.1 are included in the Hamiltonian. They are written as\cite{4,Lee}
\begin{equation}
H_0=\sum_{{\bf k}\sigma}\Psi_{k\sigma}^\dagger M_{\bf
k}\Psi_{k\sigma},
\end{equation}
where the three-component field $\Psi_{k\sigma}$ is defined as $\Psi_{k\sigma}=(
d_{xz\sigma}({\bf k}), d_{yz \sigma}({\bf k}),d_{xy\sigma}({\bf k}))^T$ and the Matrix
$M_{\bf k}$  is given by
\begin{eqnarray}
\label{Mk}
 M_{\bf
k}=\left(
\begin{array}{ccc}
\varepsilon_{11}(k)&\varepsilon_{12}(k)&\varepsilon_{13}(k)\\
\varepsilon_{21}(k) &\varepsilon_{22}(k)&\varepsilon_{23}(k)\\
\varepsilon_{31}(k)& \varepsilon_{32}(k)& \varepsilon_{33}(k)\\
\end{array}\right),
\end{eqnarray}
with elements
\begin{eqnarray}
\nonumber&&\varepsilon_{11}=-2t_1\cos k_x-2t_2\cos k_y-4t_3\cos k_x\cos k_y\\&&\nonumber
\varepsilon_{22}=-2t_2\cos k_x-2t_1\cos k_y-4t_3\cos k_x\cos k_y\\&&\nonumber
\varepsilon_{33}=-2t_7\cos k_x-2t_7\cos k_y-4t_8\cos k_x\cos k_y\\&&\nonumber
\varepsilon_{12}=\varepsilon_{21}=-4t_4\sin k_x\sin k_y\\&&\nonumber
\varepsilon_{13}=-\varepsilon_{31}=-2it_5\sin k_x\\&&\nonumber
\varepsilon_{23}=-\varepsilon_{32}=-2it_6\sin k_y.\nonumber
\end{eqnarray}
The hopping parameters are set to  $t_1 = -1.0(\approx0.4 eV), t_2 = 0.7, t_3 = -0.80,
t_4 = 0.6, t_5=t_6=-0.35, t_7=-0.3, t_8=0.2, \mu = 1.15$, in units of $|t_1|$ \cite{4}.

\begin{figure}
\label{energy}
\includegraphics[width=8cm,height=7cm]{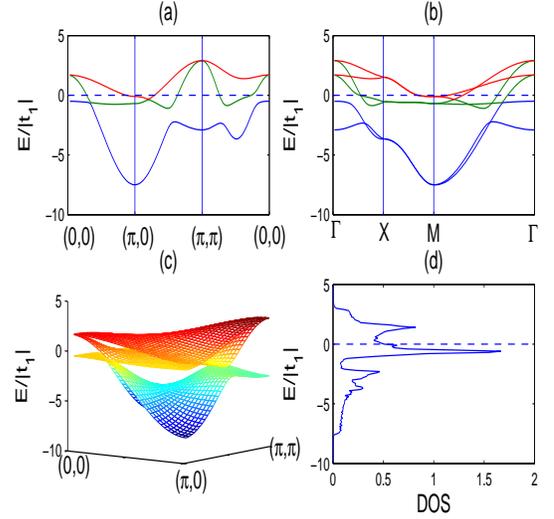}
\caption{(a) Energy dispersion in the unfolded BZ of the three-orbital model with $ t_1 =
-1.0, t_2 = 0.7, t_3 = -0.80, t_4 = 0.6,t_5=t_6=-0.35,t_7=-0.3,t_8=0.2, \mu = 1.15$ along
the path $(0,0)\rightarrow(\pi,0)\rightarrow(\pi,\pi)\rightarrow(0,0)$. (b) Energy
dispersion in the folded BZ along $\Gamma={\bf q=(0,0)},X={\bf q=({\pi\over2},
{\pi\over2})}, M={\bf q=(\pi,0)}$. (c) Energy dispersion on the $(k_x,k_y)$ plane. (d)
density of states of the three-band model, where the dashed line represents the Fermi
level.}
\end{figure}

%\begin{figure}
%\includegraphics[width=5cm,height=12cm]{Fermisur.eps}
%\caption{(a) The Fermi surface of the three-orbital model on the
%unfold BZ, with one Fe per cell. $\alpha_{1,2}$ surface are hole
%Fermi pockets and $\beta_{1,2}$ are electron Fermi pockets. The
%dashed square indicate the BZ of the two Fe per cell. (b) The Fermi
%surface folded down into the two Fe per cell BZ consists of two hole
%Fermi surface around the $\Gamma$, and two elliptically deformed
%electron Fermi surface around M point.}
%\end{figure}

%\begin{figure}
%\includegraphics[width=8cm,height=6cm]{ener1.eps}
%\caption{1}
%\end{figure}

%In this three-orbital model, the orbitals interaction between $d_{xz}$ and $d_{yz}$ like
%the two orbitals model \cite{Raghu}.
%In the x-y plane, we can find the perfect nested
%between one hole pockets around the $\Gamma$ point $({\bf q}=0,0)$ in expand Brillouin
%zone (one Fe per cell) and one Fermi pocket around the $M$ point $({\bf q}=\pi,0)$, these
%feature are supported by the result of a experiment \cite{1}. To compare with band
%structure calculations, one must fold the large BZ into a smaller one which is dual to
%the crystallographic unit cell containing two Fe atoms.
The energy dispersion is plotted in Fig. 2(a) in the extended BZ and 2(b) along the
$\Gamma\rightarrow X \rightarrow M \rightarrow\Gamma$ in the
folded BZ(two Fe atoms per unit cell). %Comparing with the two-band model, one more band is added as shown by the
%lowest line in Fig. 3(a). This band are important because it's close to the Fermi level.
Fig. 2(d) plots the density of states of the band structure. It is seen that there are
two dominating Van Hove singularities near the Fermi level, close to the hole and
electron pockets. The density of states is significantly different from the two-band
model, of which the Van Hove singularities are more distant from the Fermi level. A
stronger nesting effect is expected in this three-band model. The third band is mixed
with the two conventional bands  and thus should have great contribution to the spin
susceptibility.

%The Fermi surface is shown in Fig. 4(a) in the unfolded BZ and 4(b) in the folded BZ. It
%is seen that two hole-type FS', $\alpha_1$ and $\alpha_2$, are co-centered around the
%$\Gamma$ point, and two electron-type FS', $\beta_1$ and $\beta_2$, are ellipses around
%the M point. These Fermi surfaces are similar to those obtained from the first principle
%calculations\cite{Boeri,Haule} and those in the two-band model \cite{Raghu}.
%Preliminary data by angle resolved photoemission spectroscopy (APRES) on these crystals
%show two groups of superconducting gaps ($\Delta_1\approx12mev,\Delta_2\approx6mev$) all
%with s-wave symmetry \cite{1}, \cite{Sin}.

\section{The spin susceptibility}

The static spin susceptibility in the non-interacting case is given by $\chi^{(0)}({\bf
q}) = \sum_{ll'}\chi^0_{ll'}({\bf q})$ with
\begin{eqnarray}
\label{chill} \chi^0_{ll'}({\bf q})&={1\over 2N}\sum_{k}{f(\varepsilon_{{\bf
k}l})-f(\varepsilon_{{\bf k+q}l'})\over \varepsilon_{{\bf k+q}l'}-\varepsilon_{{\bf
k}l}}|<{k+q,l'}|{k,l}>|^2
\end{eqnarray}
where $l,l'=1,2,3$ are band indexes, $f(\varepsilon)=1/(e^{\beta(\varepsilon-\mu)}+1)$ is
the Fermi distribution function, $\beta=1/kT$, and $\varepsilon_{{\bf k}l}$ and $|{\bf
k},l> $ are the eigenvalues and eigenvectors respectively of the Hamiltonian matrix
(\ref{Mk}).

We fix $kT = 0.02(\sim 93K)$ for a finite lattice with 64 $\times$ 64 ${\bf k}$ meshes in
the extended BZ to calculate the static spin susceptibility, which is shown in
Fig.3(upper). The static spin susceptibility shows great peaks at ${(0,\pm\pi)}$ and
${(\pm\pi,0)}$. This indicates that a collinearly-striped antiferromagnetic (AFM) order
phase, a spin density wave (SDW), exhibits in oxypnictide. This feature is in agreement
with the neutron scattering measurements\cite{Ma, McGuire,T}. The SDW comes from the
strong nesting between the hole pocket at the $\Gamma$ point and the electron pocket at
the M point in the extended BZ. It is observed experimentally that this SDW peak is
significantly suppressed by F doping\cite{dong}. This is reasonable because a down-shift
of the Fermi level tends to reduce the size of electron pocket and enlarge the hole
pocket thus to suppress the nesting between them.

%The presence of the nesting between the hole and electron FS will induce a spin density
%wave instability,and F doping suppress the SDW instability and recovery the
%superconductivity.
%the physical spin susceptibility display two sets of dominant peaks,
%the largest value of the $\chi^0({\bf q},0)$ occurs around ${\bf
%q}=(\pi,0)$ and ${\bf q}=(0,\pi)$,
%A smaller peak of the spin susceptibility appears around the $X$ point, ${\bf
%(\pi/2,\pi/2)}$, in the BZ. %The former peaks are due to the nesting between the
%hole (electron) pockets,the nesting give rise to an effective
%pairing interaction.And the latter
%This peak is consistent with the result of neutron scattering measurements, which showed
%that LaOFeAs exhibits an antiferromagnetic long-range ordering with a moment of 0.35
%$\mu_B$ per Fe followed by a small distortion\cite{T}.

\begin{figure}
\label{chi}
\includegraphics[width=5cm,height=4cm]{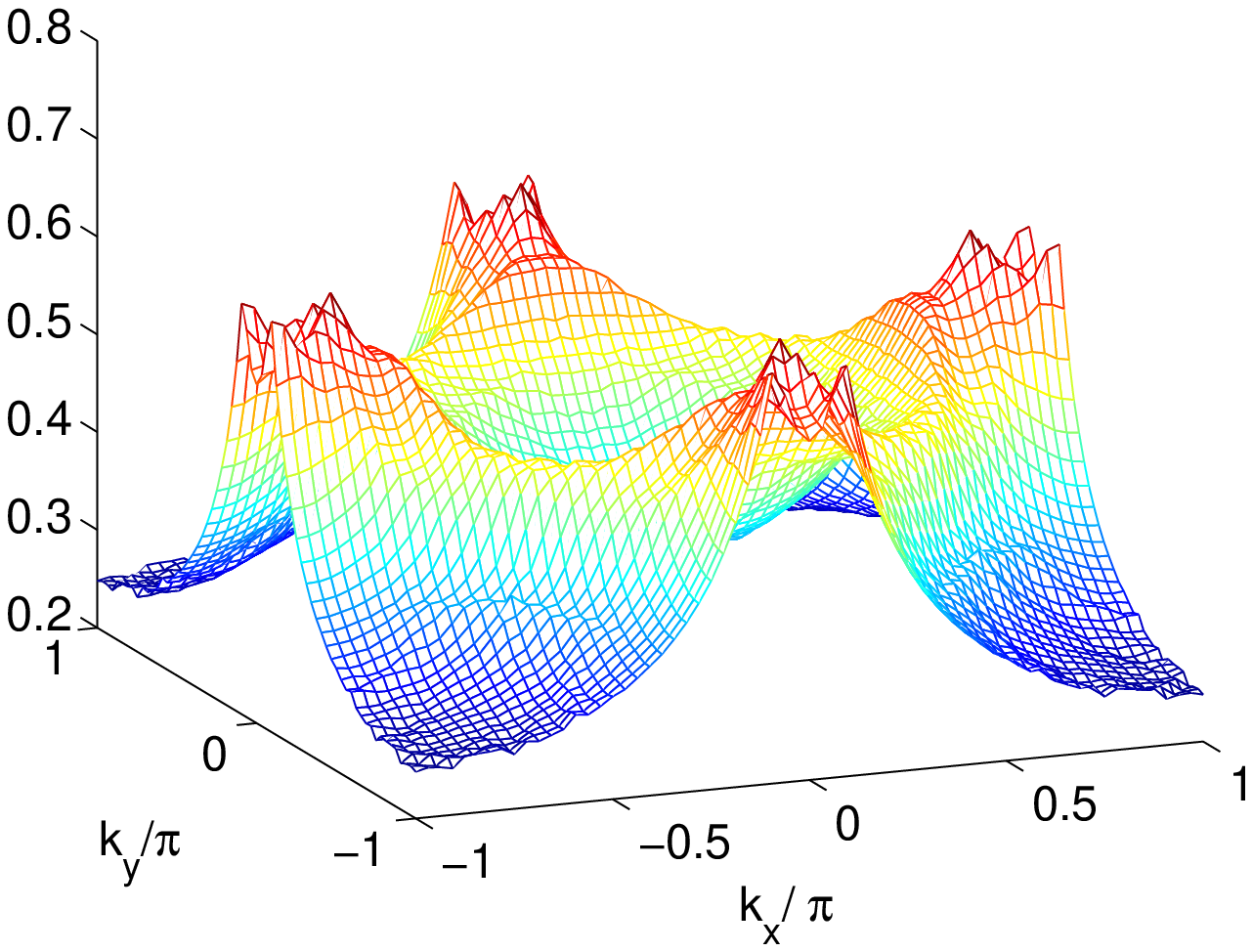}

\includegraphics[width=5cm,height=4cm]{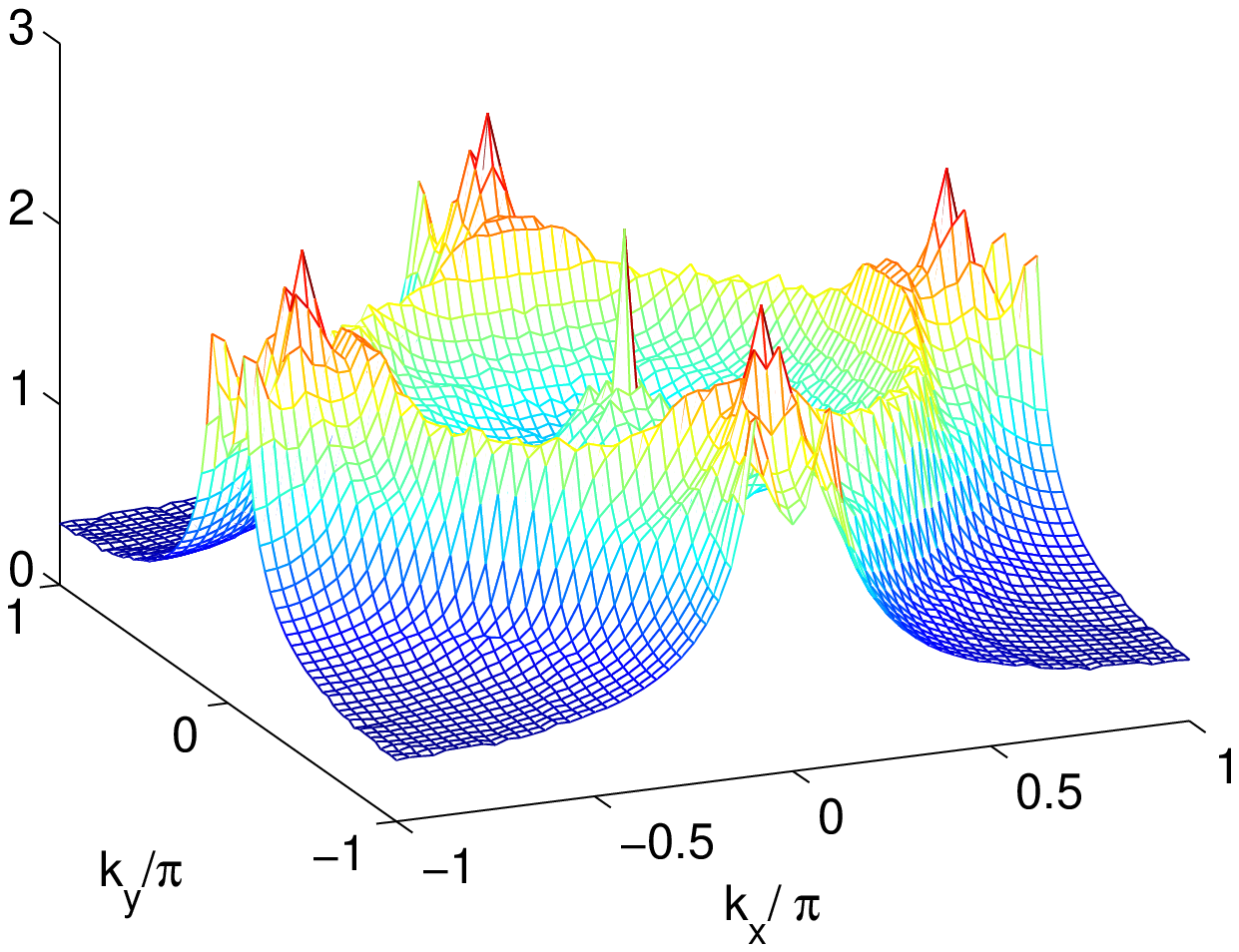}
\caption{The static spin susceptibility $\chi^{(0)}({\bf q})$ versus ${\bf q}$(upper).
The static spin susceptibility $\chi({\bf q})$ in the RPA(lower) with $U = 3.0, J=0., kT
= 0.02$.}\end{figure}
%. (c) The
%bare spin susceptibility $\chi({\bf q})$ in the whole BZ. (d) The RPA spin susceptibility
%$\chi^{RPA}({\bf q})$ for $U=1.0$ and $J=0.1$.}

%\section{Spin susceptibility and gap function in RPA}
Now we consider the interactions: the intra-band Coulomb repulsion $U$, the inter-band
coulomb interaction $U'$, the Hund's coupling $J$.  The interaction Hamiltonian is
written as
%in the first and second terms,$U$ and $U'$ are the intra- and
%inter-orbital Coulomb interactions,respectively.
\begin{eqnarray}
\nonumber H_{int}&=&U\sum_{il}n_{il\uparrow}n_{il\downarrow}+U'\sum_{i,l\neq
l'}n_{il}n_{il'}\\&+&J\sum_{i,l\neq l'}{\bf S}_{il}\cdot{\bf S}_{il'}
%+J'\sum_{il\neq
%l'}d_{il\uparrow}^\dagger d_{il\downarrow}^\dagger d_{il'\downarrow}d_{il\uparrow}
\end{eqnarray}

In the RPA\cite{2}, the spin susceptibility, $\chi^s({\bf q})$, and the charge-orbital
susceptibility, $\chi^c({\bf q})$, with interactions are given by
\begin{eqnarray}\label{chis}
\chi^s({\bf q})&&=[\hat{I}-U^s\chi^0({\bf q})]^{-1}\chi^0({\bf q}),\\
\chi^c({\bf q})&&=[\hat{I}+U^c\chi^0({\bf q})]^{-1}\chi^0({\bf q}),\label{chic}
\end{eqnarray}
%The magnetic (charge or orbital) instability appears when the
%following relation is satisfied.
%\begin{eqnarray}
%\det[\hat{I}\mp U^{s(c)}\chi^0({\bf q},\omega)]=0,\end{eqnarray} for
%$S_{ll'}({\bf q},\omega)[C_{ll'}({\bf q},\omega)]$.
%\begin{eqnarray}
%\nonumber\hat{\chi}_0({\bf
%q},\omega)=\left[\begin{array}{ccc}\chi_{11}^0({\bf q},\omega)&\chi_{12}^0({\bf
%q},\omega)&\chi_{13}^0({\bf q},\omega)\\\chi_{21}^0({\bf q},\omega)&\chi_{22}^0({\bf
%q},\omega)&\chi_{23}^0({\bf
%q},\omega)\\
%\chi_{31}^0({\bf q},\omega)&\chi_{32}^0({\bf
%q},\omega)&\chi_{33}^0({\bf q},\omega)\end{array}\right],
%\end{eqnarray}
where $\chi^0({\bf q})$ is a $3\times 3$ matrix with elements defined in (\ref{chill})
and $U^{s(c)}$ are the interaction matrices
\begin{eqnarray}
\nonumber U^s=\left[\begin{array}{ccc}U&-J&-J\\
-J&U&-J\\
-J&-J&U\end{array}\right],U^c=\left[\begin{array}{ccc}U&2U'&2U'\\
2U'&U&2U'\\
2U'&2U'&U\end{array}\right]
\end{eqnarray}

It is seen from (\ref{chis}) and (\ref{chic}) there will appears a magnetic instability
when the following relations are satisfied:
\begin{align}
det[\hat{I}\mp U^{s(c)}\chi^0({\bf q})]=0\label{det}
\end{align}
To show the magnetic instability at the M point the determinants (\ref{det}) at the SDW
momentum, $(0, \pi)$, are calculated for different interaction parameters and are plotted
in Fig.4. The parameters on the contour lines labeled by "$-0-$" in these two diagrams
give zero determinants thus lead to magnetic instabilities. For a fixed $U$ the spin
susceptibility decreases with increasing Hund's coupling $J$ thus reduces the SDW peak at
the M point. This result is quite different from that made by Raghu {\it et
al}\cite{Raghu}, who found stronger spin fluctuations with increasing $J$. The lower
diagram in Fig.3 shows the spin susceptibility close to the magnetic instability with $U
= 3.0, J = 0.0$, where the SDW peak is much enhanced due to the intraband Coulomb
repulsion relative to that of the bare spin susceptibility. When $J$ increases, however,
the SDW peak drops significantly until the SDW phase  fully disappears. It is interesting
to notice that, when $U<5$, a ferromagnetic coupling ($J<0$) is beneficial to the
formation of the SDW on the M point. On the other hand, the charge-orbital instability
appears in the region $-4.2<U'<0$. That is, when a Coulomb attraction exists between
different bands a charge-orbital instability may occur. Apart from this region the
charge-orbital susceptibility also drops rapidly. It is worthy to notice that the
charge-orbital susceptibility depends weakly on the value of $U$, the intra-band
repulsion, except for the case of very strong inter-band repulsion.

\begin{figure}
\label{detJU}
\includegraphics[width=5cm,height=4cm]{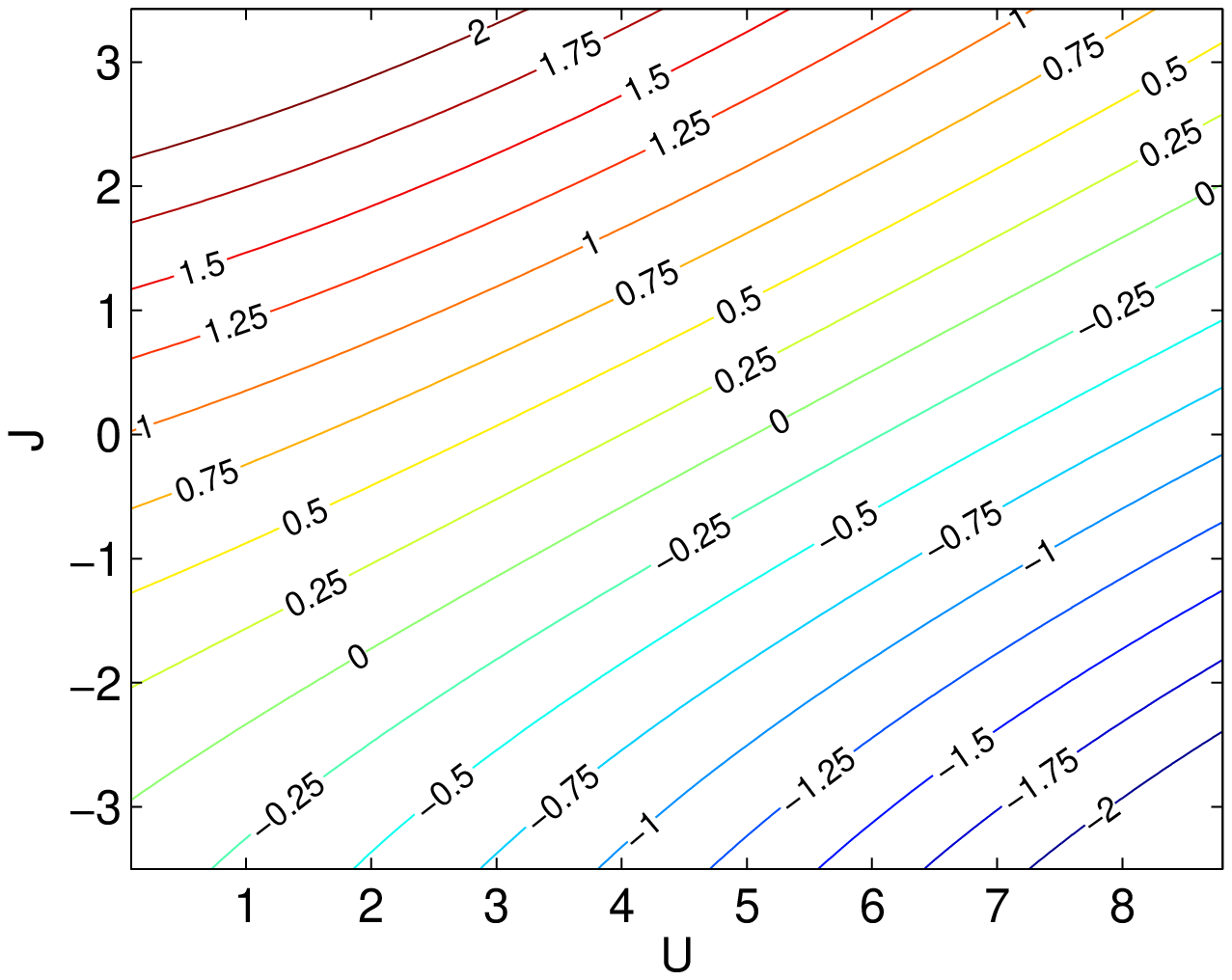}
\includegraphics[width=5cm,height=4cm]{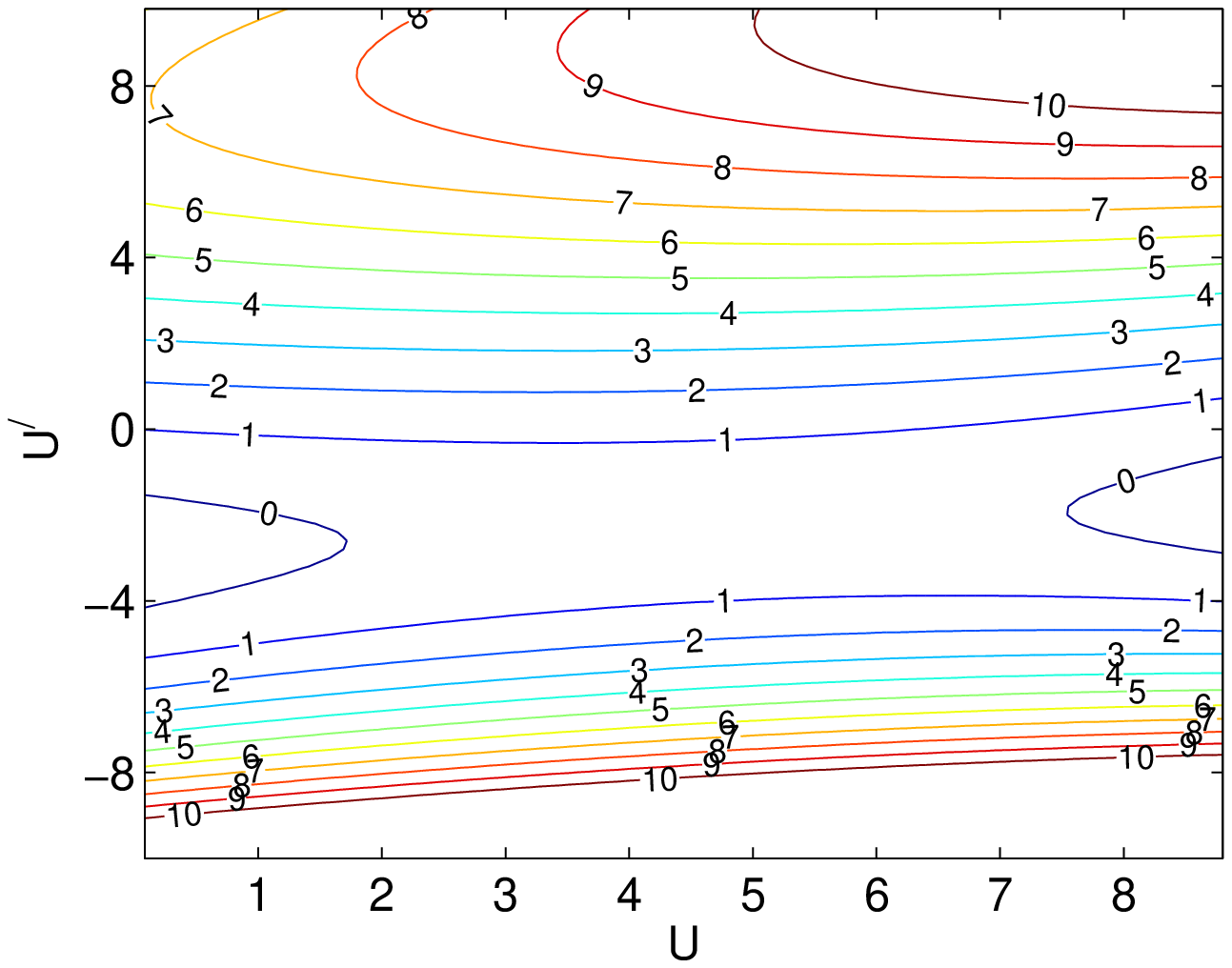}
\caption{The determinants of magnetic instabilities for $J$ vs. $U$ and $U'$ vs. $U$. The
contour line $-0-0-$ gives the parameter groups of instability. }
\end{figure}

The SDW phase has striking temperature dependence. It is found that at higher
temperatures a new resonance appears at the $\Gamma$ point, which starts to increase
rapidly at a temperature $kT \sim 0.08$, as shown in  Fig.5. This resonance corresponds
to the formation of a paramagnetic phase in the material.
%This new excitation leads to the
%formation of a paramagnetic phase in the oxypnictide material.
The SDW peak at the M point drops to a nearly stable value at this temperature. This
result is qualitatively in agreement with the experimental observation for the
stripe-like AFM phase in LaOFeAs, which forms under a temperature 134K\cite{Clarina}. At
$kT_c=0.12$ the intensity of paramagnetic phase starts to surpass that of the striped AFM
phase. It is found that temperature changes hardly the bare spin susceptibility but
smooths its distribution in the BZ. Reducing the intra-band coupling parameter $U$, the
relative intensity between the paramagnetic phase and the SDW phase reduces
significantly. Hence it is confirmed that this new resonance comes from the intra-band
coupling. %Its momentum range indicates that this intra-band coupling occurs dominantly
%inside the hole pocket, where more electrons are filled when temperature increases.
Therefore, this temperature dependence of the spin susceptibility provides detailed
information for the band structure of the oxypnictide material. More experimental
evidences for this temperature dependence are expected. At very low temperatures the spin
susceptibility showed more abundant structures but requires further stringent
calculations.

\begin{figure}
\label{tem}
\includegraphics[width=5cm,height=4cm]{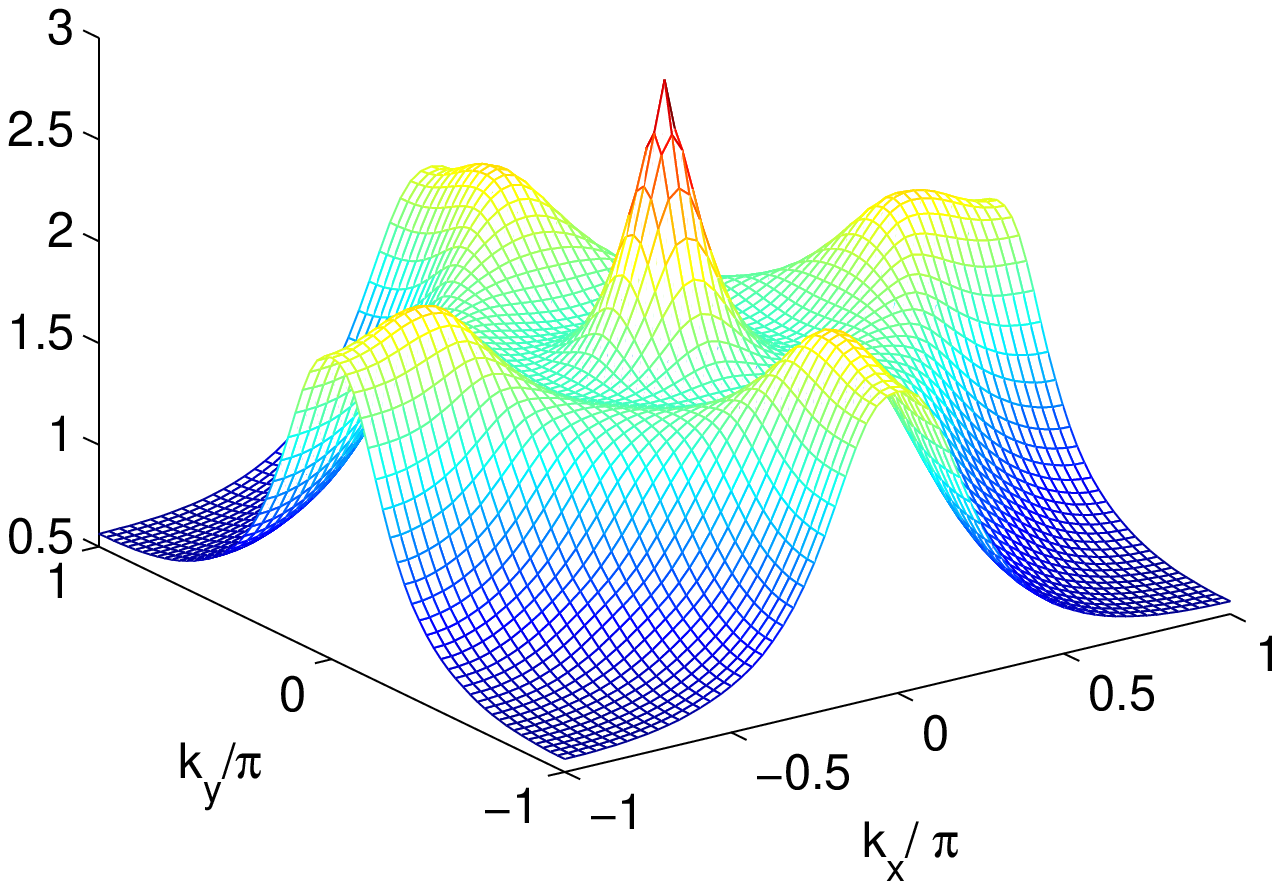}
\includegraphics[width=5cm,height=4cm]{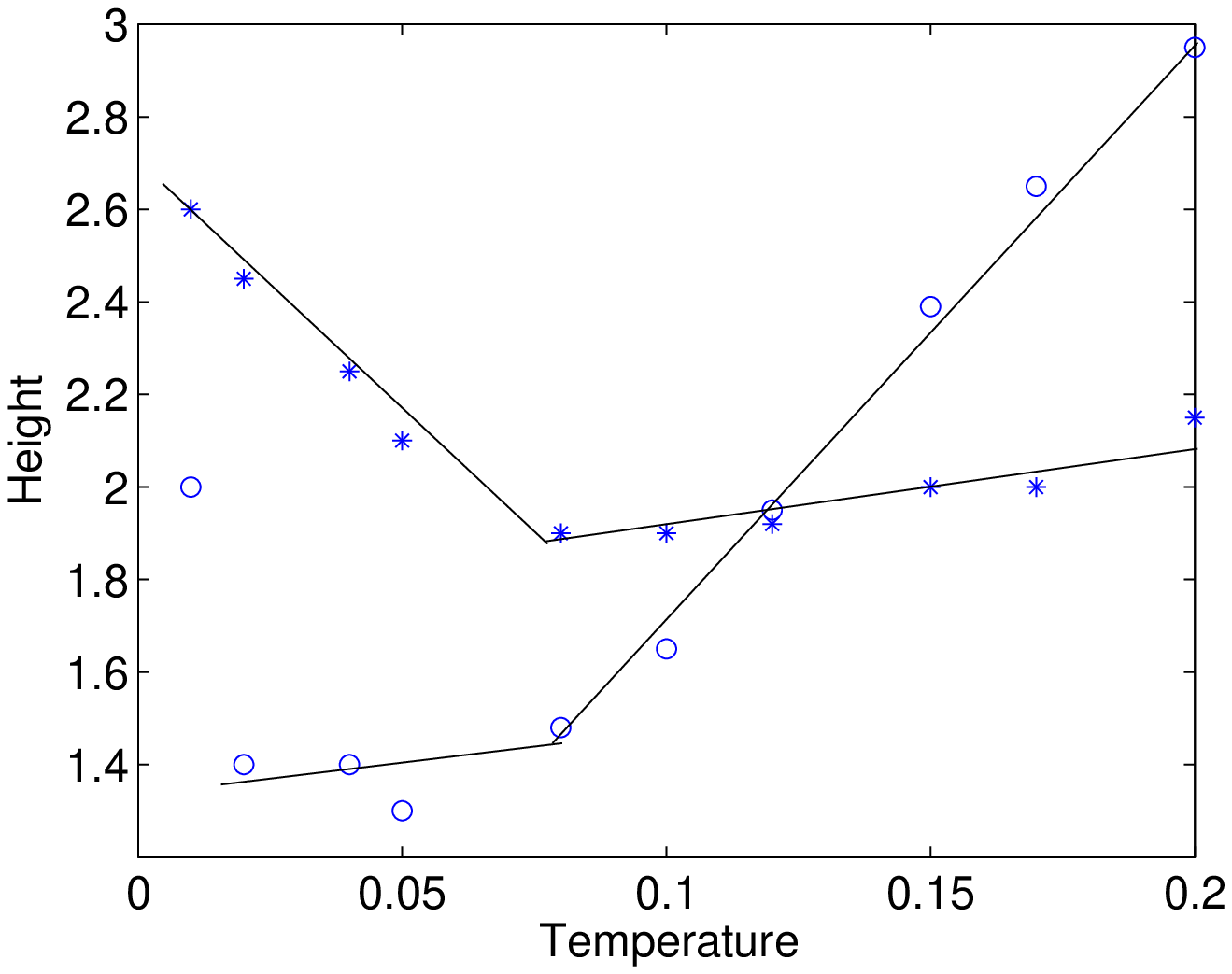}
\caption{Spin susceptibility with temperature $T=0.2$(upper) and temperature dependence
of the intensities of spin susceptibility at the $\Gamma$ point(circles) and M
point(stars) in the BZ(lower). Both results take $U=3.0, J=0, U'=1.0$. Lines are guides
for the eye.}
\end{figure}

We tried to solve the superconducting gap equation by including a pairing coupling term
into the interaction Hamiltonian but failed to find a stable gap function with some
symmetries, such as $s, s^-, d_{x^2-y^2}, d_{xy}$, etc. This failure also exists in some
literatures, e.g., Yao {\it et al} found only eigenvalues 0.1-0.4 with the $d_{xy}-$wave
symmetry\cite{Z}. But Yanagi {\it et al} claimed that some groups of interaction
parameters give eigenvalue unity for this gap symmetry in the same material\cite{YANAGI}.
Therefore, the gap symmetry remains controversy and requires further studies.

In conclusion, a three-band model is set up to reproduce the band structure near the
fermi surfaces of oxypnictide. It shows a hole pocket around the $\Gamma$ point and a
electron pocket around M point in the extended BZ. The spin susceptibility with various
interactions are calculated in the random phase approximation. It is found that the spin
susceptibility peaks around the M point, showing a SDW with momentum (0, $\pi$) and a
clear stripe-like spin arrangement. The intra-band coupling enhances remarkably the SDW
but the Hund's coupling weakens it. The interaction parameters are determined for the
magnetic instability in oxypnictide.  Finally the temperature dependence of the spin
susceptibility is studied. It is found that  a new resonance appears at higher
temperatures at the $\Gamma$ point indicating the formation of a paramagnetic phase.
There is a transition between the SDW phase and the paramagnetic phase.

%Comparity with the
%minimal two orbital model, we also can obtain the FS nesting and the
%SDW spin fluctuation peak at the hot points.Therefor,we conclude
%that this three orbital model contains a rich variety of
%magnetic,spin,orbital and pairing correlations.

This work was supported by the National Natural Science Foundation of China (Grant No.
10874049),  the State Key Program for Basic Research of China (No. 2007CB925204) and the
Natural Science Foundation of Guangdong province ( No. 07005834 ).

%\end{CJK*}

\begin{references}
%bibitem{Milman} S. A. Milman {\it et al.}, Phys. Rev. Lett. {\bf 35}, 1053 (1975).
\bibitem{Kamihara} Y. Kamihara, T. Watanabe, M. Hirano, and H. Hosonno, J. Am. Chem. Soc. {\bf 130}, 3296 (2008).
\bibitem{Chen} X. H. Chen, T. Wu, G. Wu, R. H. Liu, H. Chen, and D. F. Fang, Nature {\bf 453}, 761
 (2008).
\bibitem{dong} J. dong, H. J. Zhang, G. Xu, {\it eta al}, Europhys. Lett. {\bf 83}, 27006 (2008).
\bibitem{Mazin} I. I. Mazin, D. J. Singh, M. D. Johannes, and M. H. Du, Phys. Rev. Lett. {\bf 101},
057003 (2008).
\bibitem{Ma} C. de la Cruz, Q. Huang, J. W. Lynn, {\it eta al},  Nature {\bf 453}, 899 (2008).
\bibitem{Chen1} G. F. Chen, Z. Li, D. Wu,  {\it eta al}, Phys. Rev. Lett. {\bf 100}, 247002 (2008).
%\bibitem{2} J. Dong, H. J. Zhang, G. Xu, Z. Li,
\bibitem{McGuire} M. A. McGuire, A. D. Christianson, A. S. Sefat, {\it eta al},  arXiv:0804.0796
[cond-mat.supr-con].
\bibitem{Singh} D. Singh and M. H. Du, arXiv:0803.0429 [cond-mat.supr-con].
\bibitem{Boeri} L. Boeri, O. V. Dolgov, and A. A. Golubov, Phys. Rev. Lett. {\bf 101}, 026404 (2008).
\bibitem{Haule} K. Haule, J. H. Shim, and G. Kotliar , Phys. Rev. Lett. {\bf 100}, 226402 (2008).
\bibitem{Lee} Patrick. A. Lee, and X. G. Wen, Phys. Rev. B {\bf 78}, 144517 (2008).
\bibitem{4} S. L. Yu, J. Kang, and J. X. Li, Phys. Rev. B {\bf 79}, 064517 (2009).
\bibitem{T} T. Yildirim,  arXiv:0804.2252 [cond-mat.supr-con].
\bibitem{2} T. Takimoto, T. Hotta, and K. Ueda, Phys. Rev. B {\bf 69}, 104504 (2004).
\bibitem{Raghu} S. Raghu, X. L. Qi, C. X. Liu, D. J. Scalapino, and S. C. Zhang, Phys. Rev. B {\bf 77}, 220503(R) (2008).
\bibitem{Clarina}Clarina de la Cruz 1,2, Q. Huang3, J. W. Lynn3, arXiv:0804.0795[cond-mat.supr-con].
\bibitem{Z} Z. J. Yao, J. X. Li, and Z. D. Wang, arXiv:0804.4166 [cond-mat.supr-con].
\bibitem{YANAGI} Yuki Yanagi, Youichi Yamakawa, and Yoshiaki ONO, J. of the Phys. Soc. of Japan,
{\bf 77}, 123701 (2008).

%\bibitem{Yang} H. Yang {\it et al.}, arXiv:0803.0623 [cond-mat.supr-con].
%\bibitem{Shan} L. Shan {\it et al.}, arXiv:0803.2405 [cond-mat.supr-con].
%\bibitem{Li} Z. Li {\it et al.}, arXiv:0803.2572 [cond-mat.supr-con].

%\bibitem{Boeri} L. Boeri {\bf et al.}, cond-mat arXiv:0803. 2703 (2008 ).

%\bibitem{1} H. Ding, X. Dai, Z. Fang, G. F. Chen, J. L. Luo, {\bf et al.}, Europhys. Lett. {\bf 82} 47001 (2008).
%\bibitem{Cao} C. Cao, P. J. Hirschfeld, and H. P. Cheng, Phys. Rev. B {\bf 77}, 220506(R) (2008).

%\bibitem{Sin} D. J. Singh, and M. H. Du, Phys. Rev. Lett. {\bf 100}, 237003 (2008).


%\bibitem{C} C. de la Cruz, Q. Huang, J. W. Lynn, J. Li, W. Ratcliff II, H. A. Mook, G. F. Chen, J. L. Luo, N. L. Wang, and Pengcheng Dai, Nature, {\bf 453}, 899 (2008).
%\bibitem{F} F. Ma, and Z. Y. Lu, Phys. Rev. B {\bf 78}, 033111 (2008).
%\bibitem{Lebegue} S. Leb$\grave{e}$gue, Phys. Rev. B {\bf 75}, 035110 (2007).%ab initio calculation DFT
%\bibitem{Anderson} P. W. Anderson and W. F. Brinkman, Phys. Rev. Lett. {\bf 30}, 1108 (1973).
%\bibitem{Miyake} K. Miyake {\bf et al.}, Phys. Rev. B {\bf 34} 6554 (1986).
%\bibitem{Takimoto}T. Takimoto, Phys. Rev. B {\bf 62} R14641 (2000).
%\bibitem{Dai} X. Dai, Z. Fang, Y. Zhou and F. C. Zhang, arXiv:0803.3982 [cond-mat.supr-con].
\end{references}
\end{document}